\documentclass[10pt,letterpaper]{article}
\usepackage{ifpdf}
\ifpdf
\pdfoutput=1
\else
\fi

\usepackage{opex3}
\usepackage{amssymb}
\usepackage{amsmath}
\usepackage{cite}

\begin{document}

\title{Extremely short-length surface plasmon resonance sensors}
\author{Maxim L.~Nesterov$^{1*}$,\ Alexandre V. Kats$^{1}$ and Sergei K.~Turitsyn$%
^{2}$}


\address{$^1$A. Ya. Usikov Institute for Radiophysics and Electronics NAS of Ukraine, \\
        12 Academician Proskura Street, 61085 Kharkov, Ukraine.\\
        $^2$Photonics Research Group, School of Engineering and Applied Science, Aston University,\\
        Birmingham B4 7ET, United Kingdom.\\
        $^*$Corresponding author:}
\email{nesterovml@gmail.com} 




\begin{abstract}
The impact of the system design on the control of coupling between planar
waveguide modes and surface plasmon polaritons (SPP) is analyzed. We examine
how the efficiency of the coupling can be enhanced by an appropriate
dimensioning of a multi-layer device structure without using additional
gratings. We demonstrate that by proper design the length of the device can
be dramatically reduced through fabrication a surface plasmon resonance
sensor based on the SPP-photon transformation rather then on SPP
dissipation.
\end{abstract}




\section{Introduction}

Recent advances in the application of surface plasmon polaritons in the
fields of chemical and bio-sensors and nano-science has stimulated booming
research and development. Surface plasmon resonance (SPR) based optical
bio-sensors exploit a special type of the electromagnetic surface waves
coupled to the collective electron excitation --- surface plasmon polaritons
(see e.g. \cite{Raether,Book_Agranovich,Zayats1,Zayats_Maradudin_2005,SS,MIS}%
) to probe interactions between an analyte under study and a bio-molecular
recognition element immobilized on the SPR sensor surface. One of the most
important characteristics of the SPPs is their ability to detect small scale
effects and field interactions at the interfaces between a metal and a
non-metal. The surface plasmon polaritons attract a great deal of attention
as a quite unique possibility of the electromagnetic field localization and
corresponding substantial enhancement of the electric field near the
surface. Such a high level of a field localization results into the enhanced
sensitivity of the SPR sensors. Specifically, SPPs in sandwiched structures
have attracted much interest as a promising way to implement surface plasmon
resonance optical sensors \cite{Jiri_Homola_1999,
Jiri_Homola_2003,Paul_V_Lambeck_2006,Harris_SA_1995,Harris_BB_1999};
waveguide-coupled SPR sensor was first demonstrated by Harrie Kreuwel in  \cite{Kreuwel_ECIO_87}. Many of the key properties of
such sensors, using multi-layer sandwiched structures, are determined by the
interaction between a waveguide and SPP modes. The progress in the area is
very fast and some products are already commercially available. There is
broad range of possible schemes and configurations, and despite successful
first demonstrations of such devices, many design problems are still open
and in this paper we examine some of these issues.

We would like to point out that SPP-based technology has a very broad
spectrum of potential applications. A feasibility of exploiting the high
spectral dispersion achievable in plasmonic devices in optical communication
has been recently discussed in \cite{Telecom1}. Surface waves play an
important role in many fundamental resonant phenomena, such as e.g. the
''Wood anomalies'' in the reflectivity and transmissivity \cite
{Raether,KNN_PhysRev_2005,KNN_PhysRev_2007} of periodically corrugated metal
samples, and are already practically exploited in a wide range of practical
devices. A very important area of the SPP applications is in miniaturization
and integration of optical circuits. A possibility of high localization of
electromagnetic fields using SPP and subwavelength metal structures makes it
possible to overcome the diffraction limit of light waves which is a
critical issue in the development of photonic circuits. Considerable efforts
have been recently placed into the development of plasmon subwavelength
guiding and coupling devices\cite
{Bozh_PRL_2001,Bozh_LPL_2006,Bozh_Nature_2006,Bozh_Nature_2007}.

A promising direction in the development of \ SPP-based devices and
techniques is to combine surface plasmon approaches with well developed
dielectric waveguide technologies, in particular with fiber-optics \cite
{FO1,FO2,FO3,FO4}. The conventional modern SPR fiber sensors utilize Bragg
gratings or long period gratings to couple light to the SPPs. Recently the
coupling efficiency between the core mode of the waveguide and SPP modes of
the metal film in planar and cylindrical geometry for sensing purposes have
been examined in \cite
{KashyapOC2007planar,AlbertOL2007,Kabashin2006,KrennOE2008}. A variety of
possible configurations of multi-layered structures offers a range of
potential device designs and this field is still far from been fully
explored. Efficiency of the coupling between SPP modes and the waveguide is
the key to unlocking the full potential of such sandwiched multi-layer
structured devices. In this work we analyze the basic properties of the SPP
and waveguide mode coupling and consider a possibility to optimize the
SPR-sensors by system design through tuning device parameters rather than
through additional gratings written in the waveguide. Understanding of the
details of the interaction of the SPP and waveguide modes is a critical
point in the development of the high sensitivity real-time SPR-sensor and
our goal here is to analyze such interaction for rather general
configuration that can be relevant to many practical situations.

\section{Fields representation and boundary conditions}

\begin{figure}[h]
\centering
\includegraphics[width=8cm]{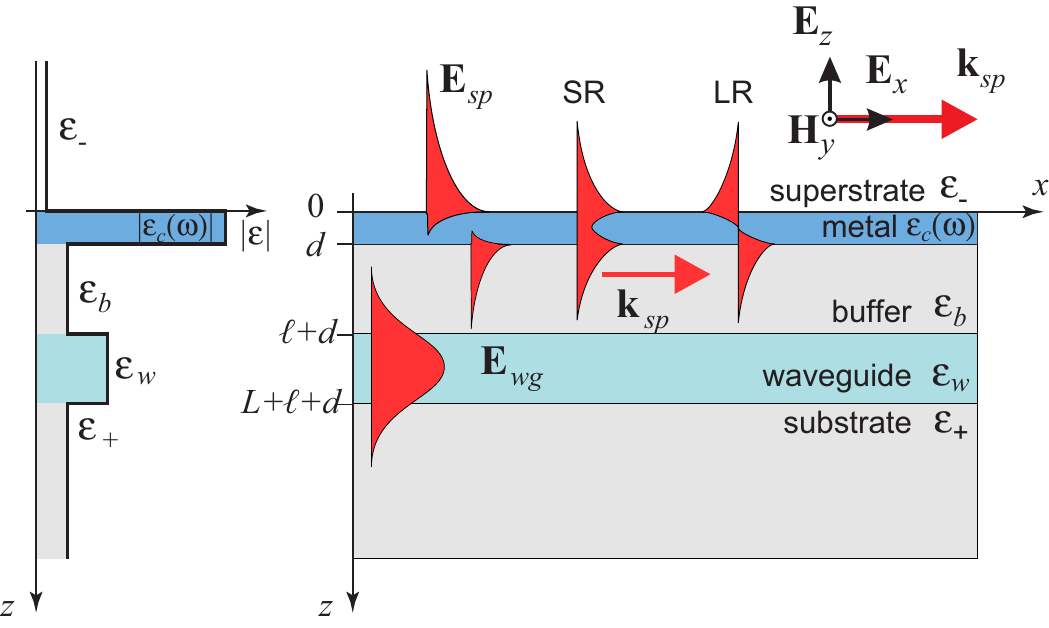}
\caption{Geometry of the problem and a typical dielectric permittivity
distribution. The structure of the SPP modes for both thick and thin metal
films are schematically shown along with a waveguide mode.}
\label{Geometry}
\end{figure}

Consider eigenmodes of the multilayer structure depicted in Fig.~\ref
{Geometry} that is typical for the SPP-based devices. The indices $c$, $b$, $%
w$, $-$ and $+$ in what follows relate to the corresponding layers as shown
in\ Fig.~\ref{Geometry}. Taking into account that SPP is a TM-polarized wave
and, therefore, in the geometry under consideration it can interact only
with TM-polarized waves, we consider TM-polarized eigenmodes ($\mathbf{E}%
=\{E_{x},0,E_{z}\}$, $\mathbf{H}=\{0,H,0\}$) of the structure. The time
dependence is moved out of the equations by the corresponding Fourier
transform and analysis of the plane waves having the form $\exp (-i\omega
t). $ We are interested here in solutions that present outgoing or decaying
waves both in the upper half-plane $z\leq 0$ ($\tau =-$), and in the region $%
z\geq L+\ell +d$ ($\tau =+$) for any given frequency $\omega $ and
tangential component of the wavevector $k_{x}$,
\begin{equation}
H^{\tau }(x,z)=H^{\tau }\exp \{ik_{x}x+i\tau k_{z}^{\tau }[z-\delta _{\tau
,+}(d+\ell +L)]\},  \label{1}
\end{equation}
where $k=\omega /c$ is the vacuum wavenumber, $k_{z}^{\tau }=\sqrt{%
k^{2}\varepsilon _{\tau }-k_{x}^{2}}$, $\mathrm{Im}(k_{z}^{\tau }),\mathrm{Re%
}(k_{z}^{\tau })\geq 0$.
Similarly,
\begin{equation}
H^{M}(x,z)=\sum_{\sigma =\pm }H^{M|\sigma }\exp \{ik_{x}x+i\sigma
k_{z}^{M}[z-z_{M}]\},  \label{2}
\end{equation}
in the metal ($M=c$, $z\in \lbrack 0,d]$), the buffer ($M=b$, $z\in \lbrack
d,d+\ell ]$) and the waveguide ($M=w$, $z\in \lbrack d+\ell ,d+\ell +L]$)
layer. Here $z_{c}=0,\,z_{b}=d,\,z_{w}=d+\ell $ and the branch of the square
roots, $k_{z}^{M}$, is chosen so that $\mathrm{Re}(k_{z}^{M})\geq 0$, $%
\mathrm{Im}(k_{z}^{M})\geq 0$ and thus the sign $\sigma =+(-)$ corresponds
to the waves propagating or decaying in the positive (negative) direction of
the $z$-axis, respectively. The $x$-dependence in all the layers is the same
due to conservation of the tangential momentum (or, equivalently, as imposed
by the boundary conditions). The electric field components can be derived
from the Maxwell equations. Assuming all the media shown in Fig.~\ref
{Geometry} to be non-magnetic we have
\begin{equation}
E_{x}^{m}(x,z)=-\frac{i}{k\varepsilon _{m}}\frac{\partial }{\partial z}%
H^{m}(x,z),  \label{22_05_08}
\end{equation}
\begin{equation}
E_{z}^{m}(x,z)=\frac{i}{k\varepsilon _{m}}\frac{\partial }{\partial x}%
H^{m}(x,z),  \label{22_05_08_1}
\end{equation}
where $m=\tau ,c,b,w$, and $\varepsilon _{m}$ denotes dielectric
permittivity in the $m$-th layer (medium), and
\begin{equation}
E_{u}^{\tau }(x,z)=E_{u}^{\tau }\exp \{ik_{x}x+i\tau k_{z}^{\tau }[z-\delta
_{\tau ,+}(d+\ell +L)]\},  \label{22_05_08_2}
\end{equation}
\begin{equation}
E_{u}^{M}(x,z)=\sum_{\sigma =\pm }E_{u}^{M|\sigma }\exp \{ik_{x}x+i\sigma
k_{z}^{M}[z-z_{M}]\}.  \label{22_05_08_3}
\end{equation}
Here $u=x,y$ stays for the corresponding components of the electric field.
The amplitudes of the electric field, $E_{u}^{\tau }$, $E_{u}^{M|\sigma }$,
can be expressed by means of Eqs.~\eqref{22_05_08}, \eqref{22_05_08_1} and %
\eqref{1}, \eqref{2} as
\begin{equation}
E_{x}^{\tau }=\frac{\tau k_{z}^{\tau }}{k\varepsilon _{\tau }}H^{\tau
},\quad E_{z}^{\tau }=-\frac{k_{x}}{k\varepsilon _{\tau }}H^{\tau },
\label{22_05_08_4}
\end{equation}
\begin{equation}
E_{x}^{M|\sigma }=\frac{\sigma k_{z}^{M}}{k\varepsilon _{M}}H^{M|\sigma
},\quad E_{z}^{M|\sigma }=-\frac{k_{x}}{k\varepsilon _{M}}H^{M|\sigma }.
\label{22_05_08_5}
\end{equation}
Introducing dimensionless variables,
\begin{equation}
\alpha =k_{x}/k,\quad \beta ^{m}=\frac{k_{z}^{m}}{k\varepsilon _{m}},
\label{22_05_08_6}
\end{equation}
Eqs.~\eqref{22_05_08_4}, \eqref{22_05_08_5} can be presented in the form
\begin{equation}
E_{x}^{\tau }=\tau \beta ^{\tau }H^{\tau },\quad E_{z}^{\tau }=-\frac{\alpha
}{\varepsilon _{\tau }}H^{\tau },  \label{22_05_08_7}
\end{equation}
\begin{equation}
E_{x}^{M|\sigma }=\sigma \beta ^{M}H^{M|\sigma },\quad E_{z}^{M|\sigma }=-%
\frac{\alpha }{\varepsilon _{M}}H^{M|\sigma }.  \label{22_05_08_8}
\end{equation}

Continuity of the tangential components of the field crossing the interfaces
results in the system of eight linear homogeneous algebraic equations for
the eight amplitudes of the magnetic field, $H^{M|\sigma }$, $H^{\tau }$.
The nontrivial solutions of these system (eigenmodes of the layered
structure) exist when the determinant is equal to zero. This leads after
straightforward algebra to the dispersion relation:
\begin{equation}
-\beta ^{w}[ia^{w}\tan \phi -b^{w}]+\beta ^{+}[a^{w}-ib^{w}\tan \phi ]=0,
\label{e2}
\end{equation}
where
\begin{equation}
\begin{split}
a^{w}& =a^{b}+b^{b}\tanh F,\quad b^{w}=\frac{\beta ^{b}}{\beta ^{w}}[%
a^{b}\tanh F+b^{b}], \\
a^{b}& =1+\frac{\beta ^{-}}{\xi }\tanh \Phi ,\quad b^{b}=\frac{\xi }{\beta
^{b}}\left[ \tanh \Phi +\frac{\beta ^{-}}{\xi }\right] ,
\end{split}
\label{e2_DOP}
\end{equation}
$\xi \equiv \beta ^{c}$ is the surface impedance of the metal film, and
\begin{equation}
\Phi =-ik_{z}^{c}d,\quad \phi =k_{z}^{w}L,\quad F=-ik_{z}^{b}\ell .
\label{e1a1}
\end{equation}
Resolving Eq.~\eqref{e2} yields the dispersion curves, $\omega =\omega
(k_{x})$ [or $k_{x}=K(\omega )$] corresponding to different modes of the
system. For the sake of clarity, but without loss of generality, in what
follows we neglect frequency dependence of all dielectric permittivities
except that one of the metal layer, $\varepsilon _{c}=\varepsilon
_{c}(\omega )$. For the latter we will use the Drude model, $\varepsilon
_{c}(\omega )=\varepsilon _{\infty }-\omega _{p}^{2}/\omega (\omega +i\gamma
)$, where $\varepsilon _{\infty }$ is high-frequency value of the
permittivity $\varepsilon _{c}(\omega )$, $\omega _{p}$ is the plasma
frequency, and $\gamma $ is the damping rate. In Table~\ref{Table} the
typical parameters for the commonly used materials (for design of the
sensors) are shown, here $n_{m}=\sqrt{\varepsilon _{m}}$, $m=-,b,w,+$,
denote refractive indices of the corresponding dielectric layers.
\begin{table*}[ht]
\caption{Table of common material parameters for the SPR sensing devices$%
^{\ast }$}
\label{Table}\centering
\begin{tabular}{rccc}
\cline{2-4}
\textbf{metal} & silver\cite{Jiri_Homola_1999} & gold\cite
{Jiri_Homola_1999,KashyapOC2007planar,AlbertOL2007} & aluminium\cite
{Nakano1994} \\ \cline{2-4}
\textbf{dielectric core} & $n=1.585$\cite{Chen} & $n=1.476$\cite{Nakano1994}
& $n=1.47$\cite{KashyapOC2007planar} \\
\textbf{dielectric cladding} & $n=1.439$\cite{Chen} & $n=1.473$\cite
{Nakano1994} & $n=1.45$\cite{KashyapOC2007planar}
\end{tabular}
\newline
\vspace{2mm} $^*$ For optical constants of silver, gold and aluminium see
\cite{Ordal,Palik,FranciscoLuis2008Const}.
\end{table*}

The field amplitudes can be expressed in terms of the upper medium magnetic
field amplitude, $H^{-}$:
\begin{equation}
\begin{split}
H^{c|\sigma }& =\frac{H^{-}}{2}\left( 1-\sigma \frac{\beta ^{-}}{\beta ^{c}}%
\right) ,\quad H^{b|\sigma }=\frac{H^{-}}{2}\{a^{b}-\sigma b^{b}\}, \\
H^{w|\sigma }& =\frac{H^{-}}{2}\{a^{w}-\sigma b^{w}\},\quad
H^{+}=H^{-}[a^{w}\cos \phi -ib^{w}\sin \phi ].
\end{split}
\label{e1a2}
\end{equation}

The dispersion relation includes a number of limiting cases corresponding to
different combinations of thick or thin limits of the building layers. For
instance, in the limit of a large thickness of the buffer layer, $\ell
\rightarrow \infty $ ($F\gg 1$), the dispersion relation splits into two
dispersion branches,
\begin{equation}
\beta ^{b}+\beta ^{+}-i\left( \beta ^{w}+\frac{\beta ^{+}\beta ^{b}}{\beta
^{w}}\right) \tan \phi =0,  \label{24_05_08}
\end{equation}
\begin{equation}
a^{b}+b^{b}=0,  \label{24_05_08_1}
\end{equation}
which correspond to the waveguide TM modes and SPP modes of the metal film,
respectively. In turn, the derived SPP dispersion relation, Eq.~%
\eqref{24_05_08_1}, for a rather thick metal film ($\Phi \gg 1$) limit leads
to standard SPP modes, $a^{b}+b^{b}=0\Rightarrow (\beta ^{b}+\xi )(\beta
^{-}+\xi )=0$. In the case of a finite value of $\Phi $ the dispersion
relation describes the hybridized double-interface-localized SPP modes,
\begin{equation}
(\beta ^{b}\tanh \Phi +\xi )(\beta ^{-}\tanh \Phi +\xi )-\xi ^{2}\cosh
^{-2}\Phi =0.  \label{e5}
\end{equation}
More specifically, when the dielectric permittivities of the superstrate and
the buffer coincide, $\varepsilon _{b}=\varepsilon _{-}$, then Eq.~\eqref{e5}
splits into two branches and yields the dispersion relations for the
short-range (symmetric) SPP and long-range (antisymmetric) SPP modes\cite
{Raether}, see Fig.~\ref{Geometry},
\begin{equation}
\beta ^{b}=\beta _{+}^{b}\equiv -\xi \tanh (\Phi /2),\quad \beta ^{b}=\beta
_{-}^{b}\equiv -\xi \coth (\Phi /2),  \label{e6}
\end{equation}
where subscripts ``$+$'' and ``$-$'' correspond to the long-range (LR), and
short-range (SR) mode, respectively.

Similar to that, the solitary waveguide modes ($\ell \rightarrow \infty $)
depend on the waveguide layer thickness and permittivity. Namely, in this
limiting case the solution of Eq.~\eqref{24_05_08} reads
\begin{equation}
\varepsilon _{w}kL\!=\!\frac{1}{\beta ^{w}}\arctan \!\! \left[ -i\frac{%
(\beta ^{b}+\beta ^{+})\beta ^{w}}{(\beta ^{w})^{2}+\beta ^{b}\beta ^{+}}%
\right]\!\! +\frac{n\pi }{\beta ^{w}}, \quad n=0,\pm 1,\ldots  \label{e10}
\end{equation}

At the crossing points of the dispersion curves the phase matching
conditions are satisfied and an interaction between waves of different
branches is most efficient. However, in order to stress the difference
between the situations when the phase matching conditions are provided by an
additional grating and the problem considered here we will use in what
follows term hybridization of modes. In the problem considered here the
finite thickness of the buffer layer results in hybridization (combination)
of the waveguide and SPP modes. The resulting hybrid eigenmodes of the
system might have spatial distributions with energy both in the dielectric
waveguide and in the SPP modes. We would like to emphasize that the most
interesting situations for the design of the sensors takes place when the
level of the mode hybridization (strong coupling of the waveguide and SPP
modes) is rather high. This occurs when the buffer layer is relatively thin
and in the vicinity of the points in the $\omega -k_{x}$ plane where the
dispersion curves of the free SPP and waveguide modes intersect. In the
vicinity of such points the hybridization is high and control of operation
near this point can be provided by the adjustment of the buffer layer. This
consideration outlines the design rules and explains the choice of the
parameters in the examples presented below.

\begin{figure*}[!bt]
\centering
\includegraphics[width=13cm]{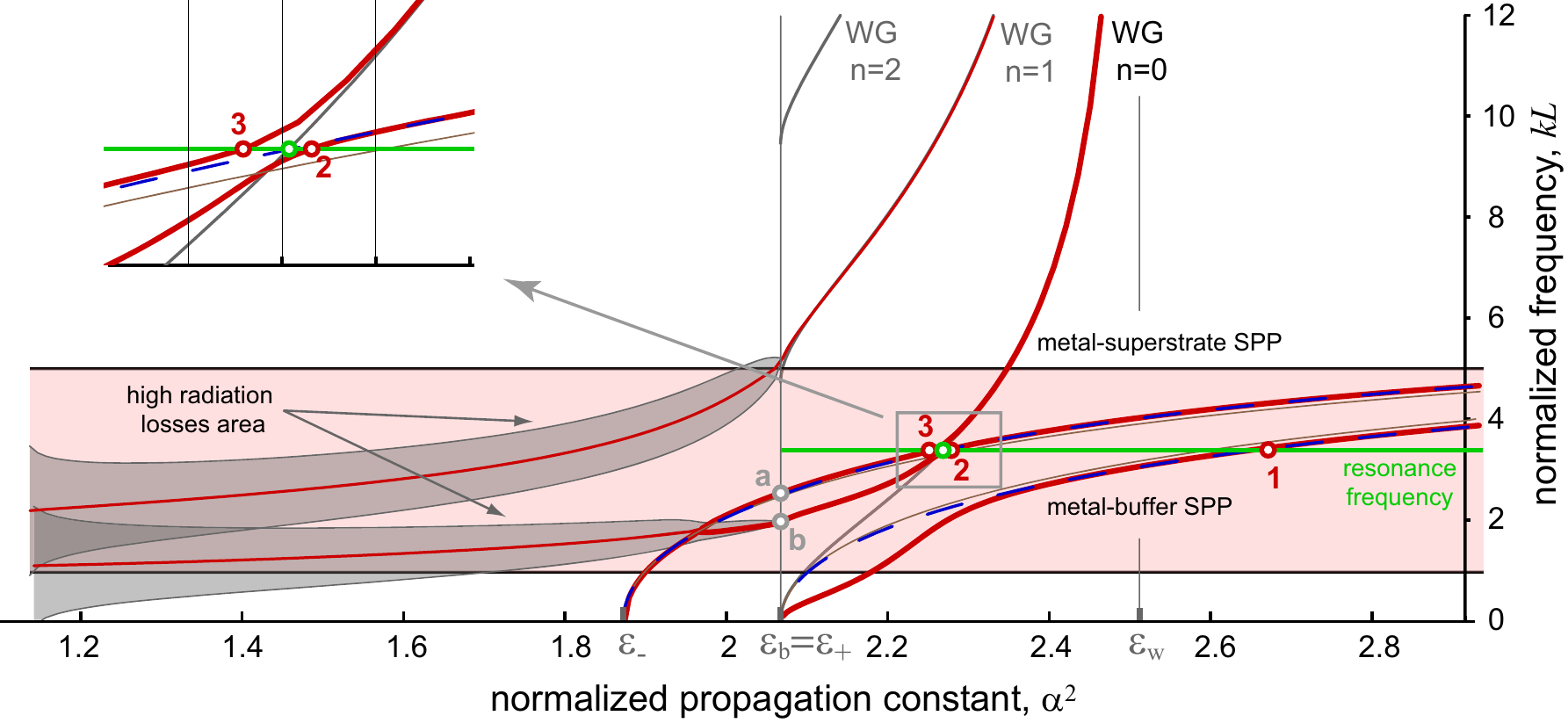}
\caption{Dispersion relations for the multi-layer system of Fig.~\ref
{Geometry}. Thick (red) curves represent the solutions of the complete
dispersion relation given in Eq.~\eqref{e2}. Thin (gray) WG-named curves
correspond to the dispersion relation of the solitary dielectric waveguide,
Eq.~\eqref{e10}, and a thick silver metal film, $d\rightarrow\infty$. The
dashed (blue) curves represent the high-frequency and low-frequency
solutions of the dispersion relation for a solitary thin silver metal film.
The parameters are: $n_-=1.37$, $n_b=n_+=1.439$, $n_w=1.585$, the
thicknesses of the layers are $d=58$~nm, $\ell=400$~nm, $L=250$~nm.}
\label{DispR}
\end{figure*}

The solutions of the dispersion relation of the multi-layer system, Eq.~%
\eqref{e2}, are plotted in Fig.~\ref{DispR} where we use the normalized
frequency parameter $kL=L\omega /c$, versus the squared normalized
tangential component of the wavenumber (i.e., the squared effective
refraction index, $n_{\mathrm{eff}}^{2}\equiv \alpha
^{2}=c^{2}k_{x}^{2}/\omega ^{2}$). The most interesting area for the
parameters lies is the vicinity of the intersection between the
silver-superstrate SPP branch and that of the lowest-frequency (fundamental)
waveguide mode, see inset in Fig.~\ref{DispR}.

The coupling of these modes due to the finite-thickness of the buffer layer
results in a repulsion of the dispersion branches. However, in the vicinity
of the inoculating curves intersection the distance between the modified
dispersion branches is relatively small. The intersection defines the
``resonance frequency'', $k_{\mathrm{res}}L=3.37$ ($\omega _{\mathrm{res}%
}/(2\pi )=6.44\cdot 10^{14}$~Hz) for $L=250$~nm, see Fig.~\ref{DispR}, that
optimizes the considered structure properties for sensor applications. Note
that our approach can be also used in the reverse problem - when the design
of the multi-layer structure is adjusted to some required operational
frequency.

\begin{figure}[!htb]
\centering
\includegraphics[width=9cm]{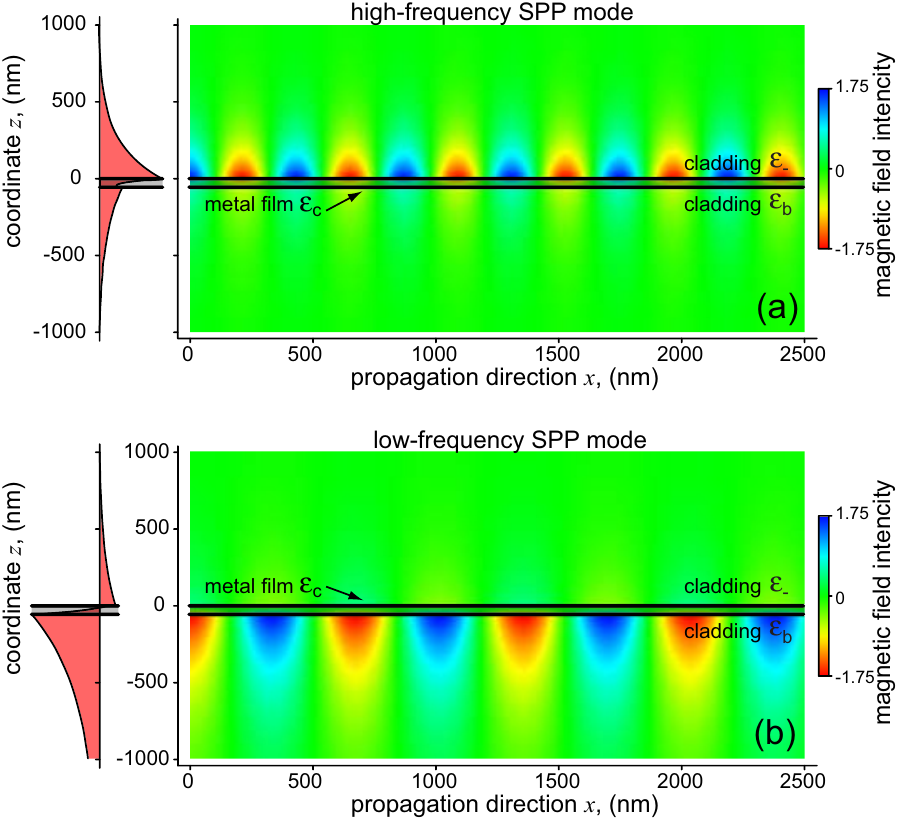}\newline
\caption{The magnetic field distribution for the two SPP eigenmodes of the
solitary metal film at $\protect\omega =\protect\omega _{\mathrm{res}}$ ($%
\protect\omega_{\mathrm{res}}L/c=3.37$). $\protect\alpha _{\mathrm{high}%
}^{2}=2.268$, $\protect\alpha _{\mathrm{low}}^{2}=2.649$ other parameters
are the same as in Fig.~\ref{DispR}. The a (b) figure is the field
distribution for the high-frequency (low-frequency) SPP mode of the metal
film corresponding to the upper (lower) SPP dispersion curve in Fig.~\ref
{DispR}.}
\label{FilmFieldSym}
\end{figure}

The magnetic field evolution along the $x$-axis for the solitary metal film
SPP modes, i.e., when the buffer and the waveguide thicknesses are
zero-valued, $\ell ,L\rightarrow 0$, could be derived from Eqs.~\eqref{1},~%
\eqref{2} and~\eqref{e1a2}, where $k_{x}$ and $\omega $ obey Eq.~\eqref{e5}.
This distribution in $(z,x)$ plane is shown in Fig.~\ref{FilmFieldSym} for
the above ``resonance frequency'', $\omega _{\mathrm{res}}$.

\begin{figure}[h]
\centering
\includegraphics[width=9cm]{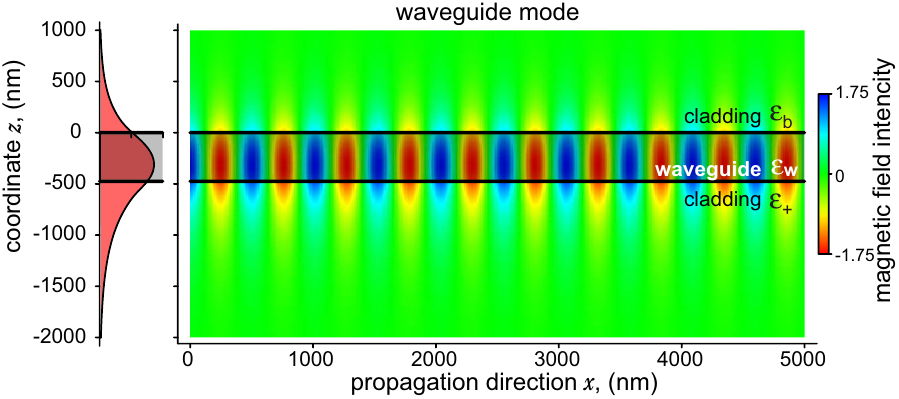}\newline
\caption{The magnetic field evolution for the solitary waveguide mode, $%
\protect\alpha =1.56$, other parameters are the same as in Fig.~\ref{DispR}}
\label{wgfield}
\end{figure}

Field evolution for the eigenmodes of the solitary dielectric waveguide (we
consider here the limit or vanishing buffer and the metal layer thicknesses:
$d,\,\ell \rightarrow 0$) could be derived from the same equations but for $%
k_x$ and $\omega $ obeying Eq.~\eqref{e10}. The corresponding magnetic field
distribution is shown in Fig.~\ref{wgfield}, for $\omega=\omega_{\mathrm{res}%
}$.

\begin{figure}[!htb]
\centering
\includegraphics[width=9cm]{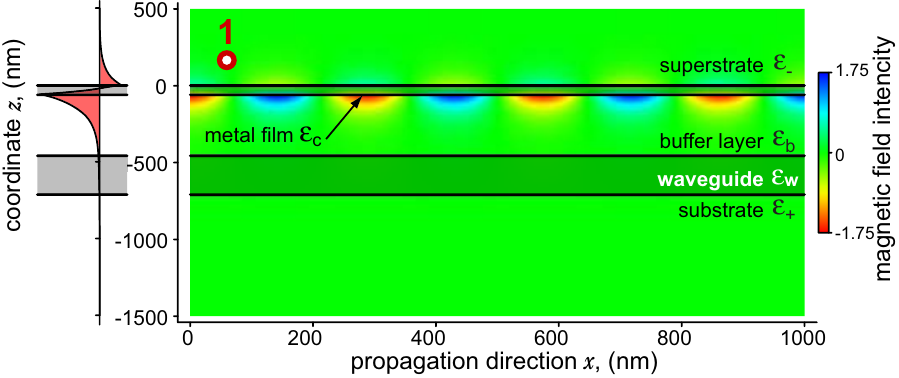}\newline
\caption{The magnetic field evolution for the eigenmode of the system
corresponding to the point 1 in Fig.~\ref{DispR}. Compare the field
distribution with that of the right in Fig.~\ref{FilmFieldSym}. The
parameters are the same as in Fig.~\ref{DispR}. }
\label{SystemFieldSym_p3}
\end{figure}

\begin{figure}[!htb]
\centering
\includegraphics[width=9cm]{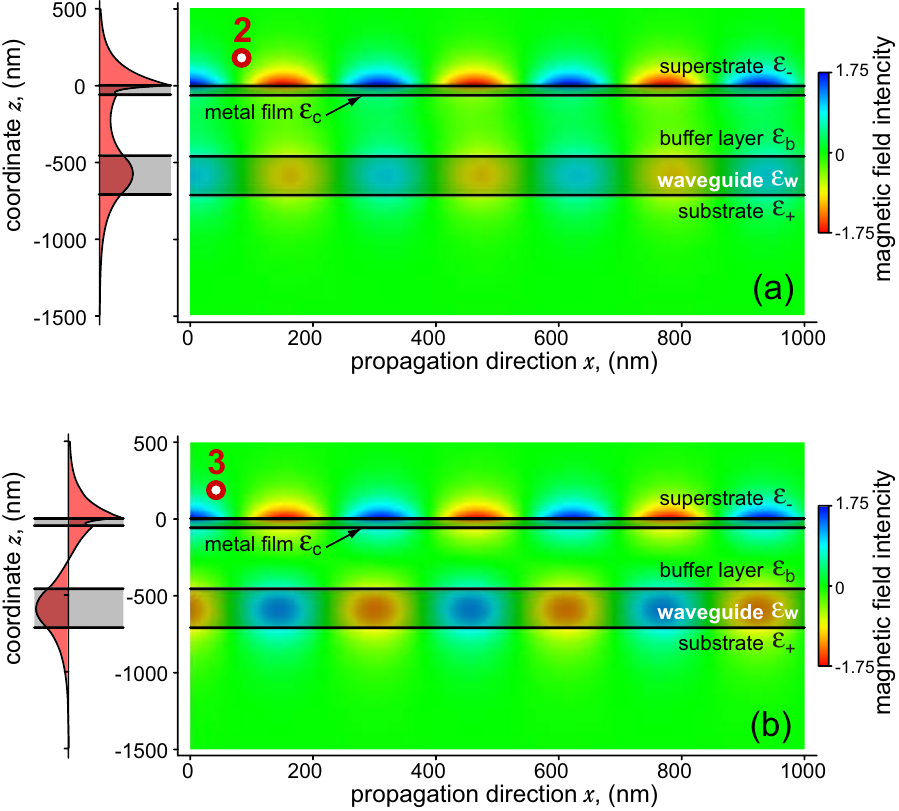}\newline
\caption{The magnetic field evolution for the eigenmodes of the system
corresponding to the points 2 (a) and 3 (b) in Fig.~\ref{DispR}.}
\label{SystemFieldSym}
\end{figure}

\begin{figure}[!h]
\centering
\includegraphics[width=9cm]{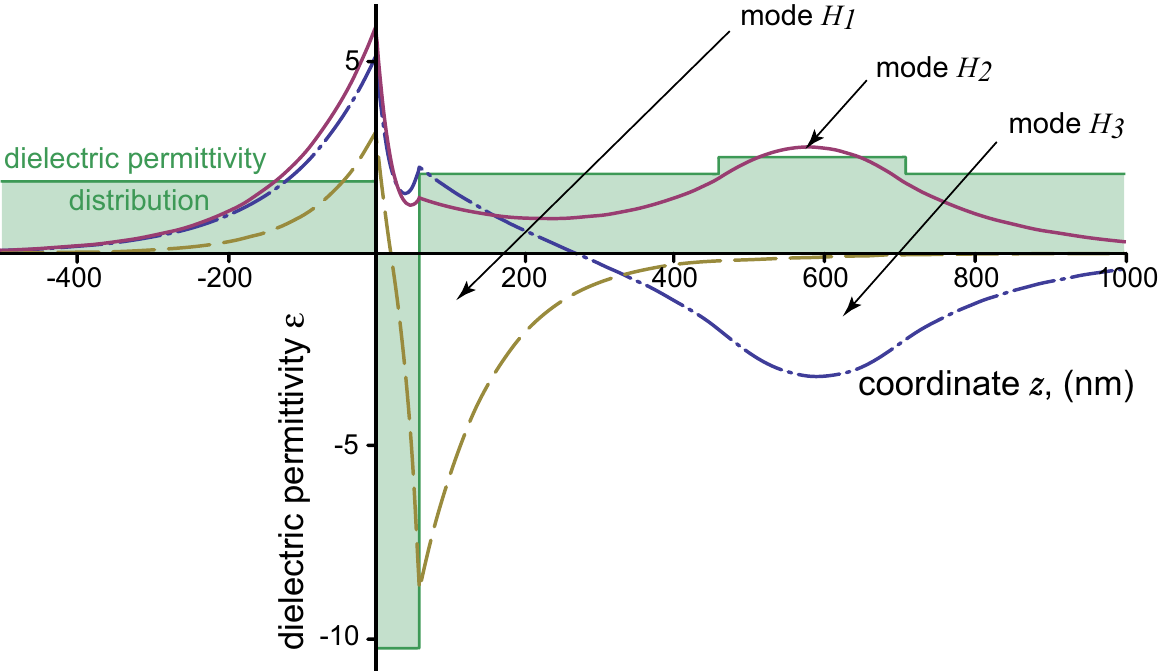}\newline
\caption{Squared magnetic field evolution for three eigenmodes of the system
for the points 1, 2 and 3 of Fig.~\ref{DispR}, and the dielectric
permittivity distribution. The parameters are the same as in Fig.~\ref{DispR}%
. }
\label{FullModes}
\end{figure}

The magnetic field distributions in the plane $(z,x)$ for three eigenmodes
of the multilayer system at the ``resonance frequency'' are shown in Figs.~%
\ref{SystemFieldSym_p3},~\ref{SystemFieldSym} and \ref{FullModes}.

It is seen that by an appropriate fitting of the parameters of a multi-layer
structure it is possible to switch between different regimes and to control
efficiency of coupling between the waveguide mode and SPP modes.

\section{Transmission properties of the multi-layer structure}

Consider now transmittance of the described above multi-layer system with
the length of the metal layer that differs from the dielectric waveguide
length. We calculate here the transmittance as a ratio between the total
output energy flux at $x=100~\mathrm{\mu m}$, and the input energy flux at $%
x=0$. The input energy flux at the starting point $x=0$ corresponds to the
solitary waveguide fundamental mode, and the output energy flux is obtained
by integrating the $x$-component of the energy flux density across the
section $x=100~\mathrm{\mu m}$ from $z=-3~\mu \mathrm{m}$ to $z=6~\mu
\mathrm{m}$. The electromagnetic field within the structure was obtained by
direct numerical simulations of the problem by means of the finite element
method (FEM). The calculation length of 100~$\mu \mathrm{m}$ was chosen
because it was found to be enough to observe the effect. The computation of
the longer structures requires significantly larger computer resources.
However, it does not add significantly new information --- use of a longer
computational domain leads to a more precise characterization of a fine
structure of the region near the minimum at the resonance frequency (it
becomes slightly wider and deeper), but the main features of the frequency
dependence of the transmittance are practically not changed.

\begin{figure}[tbh]
\centering
\includegraphics[width=9cm]{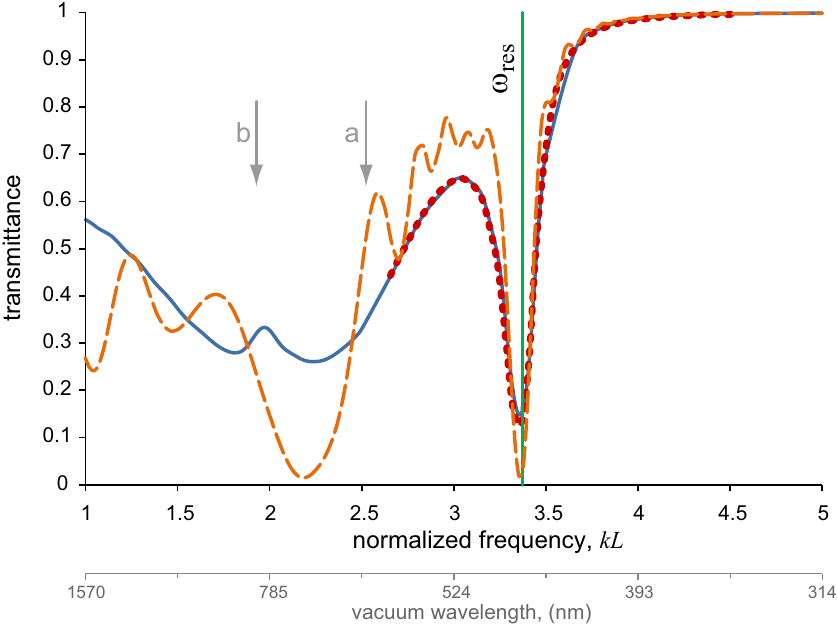}\newline
\caption{Frequency dependence of the transmission coefficient for the metal
film length $\mathcal{L}>100~\mathrm{\protect\mu m}$ (solid line), and for $%
\mathcal{L}=40~\protect\mu \mathrm{m}\sim X_{\mathrm{beat}}/2$ (dashed line)
obtained by direct numeric simulations. The ripples are caused by the finite
film length. The arrows $a$ and $b$ indicate the frequency values below
which the radiative losses arise for the modes 3 and 2, respectively. Dotted
lines present results obtained by the mode expansion procedure. The
thickness of the buffer layer is $\ell =400$ nm.}
\label{tOmega2a}
\end{figure}

The results of the numerical modeling of the transmission coefficient for
the long metal film length $\mathcal{L}>100~\mu \mathrm{m}$ (formally, $%
\mathcal{L}\rightarrow \infty $) are presented in Fig.~\ref{tOmega2a}. The
frequency range in the calculations corresponds to the stripe area in the
Fig.~\ref{DispR}.

\begin{figure}[tbh]
\centering
\includegraphics[width=13cm]{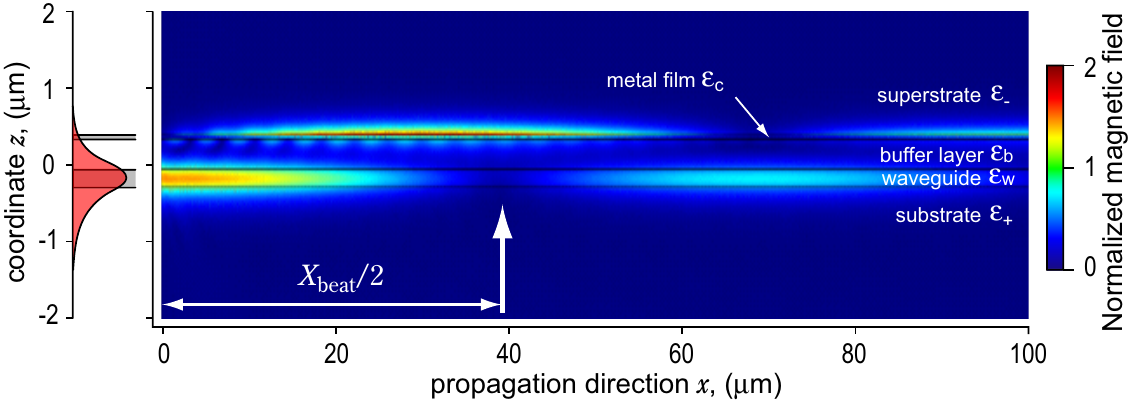}\newline
\caption{The squared magnetic field distribution at the resonance frequency,
direct numeric simulation.}
\label{field1}
\end{figure}

One can see rather narrow dip in the transmission coefficient near the
defined above ``resonance frequency'', shown by the vertical line.
Evidently, the transmission falls due to excitation of the SPP modes
accompanied by strong enhancement of the dissipative losses. Other local
minima of the transmission coefficient are caused by appearance of the
radiation losses, cf.~Fig.~\ref{DispR}. We would like to emphasize that in
the considered structure there exists 3 eigenmodes of the system
simultaneously (for some given input frequency), see Fig.~\ref{DispR}. This
fact leads to the field interference pattern easily seen in Fig.~\ref{field1}%
. For the ``resonance frequency'' the largest period of the pattern is given
by the following simple expression:
\begin{equation}
X_{\mathrm{beat}}=\frac{2\pi }{k_{x2}-k_{x3}},  \label{length1}
\end{equation}
where $k_{x2}$ and $k_{x3}$ are the real parts of the propagation constants
of the long-period modes 2 and 3 of the complete system, and could be found
from the dispersion relation. In this particular instance $X_{\mathrm{beat}%
}\approx 69.5~\mu \mathrm{m}$. The small features in the pattern are due to
interference of the short-period mode 1 with the long-period modes 2 and 3.
That is, corresponding characteristic length of the pattern is $X_{\mathrm{%
beat}}^{\prime }\simeq 2\pi /(k_{x1}-k_{x2})\simeq 3.9~\mu \mathrm{m}$, cf.
Fig.~\ref{field1}. These features are in excellent agreement with the
results of the recent experimental work \cite{KrennOE2008}, where authors
consider similar geometry and discuss the coupling length and energy
exchange between waveguide and SPP modes.

At the input of the system main part of the energy was in the dielectric
waveguide and then is gradually transferred to the metal film. It is seen
that the beat half-period $X_{\mathrm{beat}}/2$ gives us a characteristic
length for energy exchange between the core- and the high-frequency metal
film-localized mode (coupling length) and at the distance $x=X_{\mathrm{beat}%
}/2$ the maximal transfer of the energy into the metal-film-localized mode
occurs, see Fig.~\ref{field1}. Consequently, low transmission coefficients
(corresponding to almost total energy transfer from waveguide to the SPP)
can be achieved by choosing the metal film length of the order $\mathcal{L}_{%
\mathrm{opt}}=X_{\mathrm{beat}}/2$, see Fig.~\ref{tOmega2a}. This distance
is defined as a minimal length of the metal film that can provide for the
total energy transfer from the waveguide core to the metal. Since the
process of the energy transfer from the core to the SPP and in the backward
direction in the absence of losses is cyclic, there exist a number of metal
film lengths for which the energy total vanishing in the core appears: $%
\mathcal{L}_{m}=X_{\mathrm{beat}}/2+mX_{\mathrm{beat}}$; $m=0,1,2,...$.
However, the minimal metal film length has advantage of a large contrast
between resonant and non-resonant transmission (transmission at resonant and
non-resonant frequencies) due to the small dissipation losses in the shorter
metal film. Additionally, the SPP propagation length in our geometry at the
resonance frequency is about 100 $\mu $m, that is longer than the minimal
length of the metal film. In other words we propose here to make a SPR
sensing device, that is shorter (comparable) than SPP propagation length, in
contrast to the commonly studied devices with metal film that is much longer
than SPP propagation length \cite
{Harris_SA_1995,Harris_BB_1999,Ctyroky_SA_1999}. We anticipate that devices
with such a minimal length might also have enhanced sensitivity. In the
recent work \cite{Dwivedi_2008} it was observed that with decrease in
sensing region length, the sensitivity, signal-to-noise ratio are both
increasing and the resolution also improves. However, analysis of the
sensitivity and signal-to-noise ratio issues for the devices with minimal
sensing region length proposed here are beyond the scope of this paper and
will be discussed elsewhere.

\begin{figure}[tbh]
\centering
\includegraphics[width=13cm]{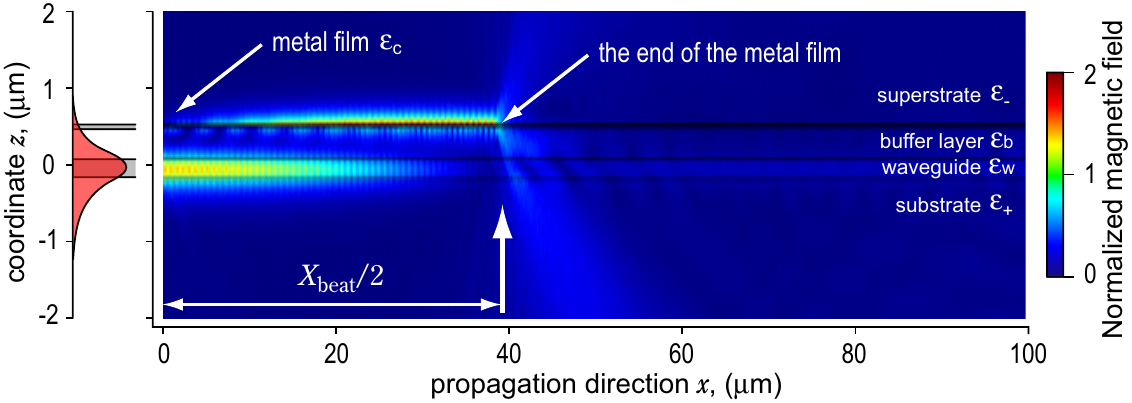}\newline
\caption{The squared magnetic field evolution at the resonance frequency for
the structure with the metal film, which is cut at some special length of 40~%
$\protect\mu \mathrm{m}$. Direct numeric simulation.}
\label{field2}
\end{figure}

We stress that it is clearly seen in Fig.~\ref{tOmega2a} that the effect is
even more pronounced as compared with the long-film system. Indeed, the
transmittance is about zero at the resonance frequency for the dashed curve
corresponding to the metal film length of $\mathcal{L}\thickapprox X_{%
\mathrm{beat}}/2$. This means, in particular, that there is no need to make
a millimeter-long \cite{Harris_SA_1995,Harris_BB_1999,Ctyroky_SA_1999} film
in order to observe strong transmittance suppression due to the SPR
resonance. This feature might be of a particular interest for device
miniaturization and integration. The ripples in the frequency dependence of
the transmittance coefficient are due to the finite length of the metal
film. In Fig.~\ref{field2}
the distribution of the squared magnetic field is depicted. It is simulated
at the resonance frequency for the system with the metal film of the length $%
\mathcal{L}=40~\mu \mathrm{m}$. It is seen that, indeed, near the metal film
end the energy flux is concentrated close to the film, demonstrating
approximately a total energy transfer from the core into the SPP mode and
consecutive scattering out of the system at the metal film edge. As a
result, the energy flux in the cross-section $x=100$~$\mu $m is practically
disintegrated.

Finally, we would like to discuss briefly the possibility to apply
approximate approach that can be used when a direct numerical modeling
becomes time consuming. The fields evolution and transmittance through the
structure can be approximately found using the well-known mode expansion
method \cite{Yariv}. We would like to point out that there exist rather
advanced theoretical methods for analysis of the  excitation of surface
plasma waves in integrated-optical waveguide structures \cite
{Ctyroky_SA_1999}, however, in our geometry it appears to be sufficient to
apply a more simple and straightforward mode expansion approach in the
following realization. Note that in the considered configurations the
homogeneity of the multi-layer structure along the $x$-axis is preserved
(e.g. no gratings are used), there will be no energy transfer (energy
exchange) between the eigenmodes of the complete structure during the
propagation. Therefore, it is possible to express any arbitrary incoming
field distribution --- the fundamental waveguide mode of the dielectric
waveguide $H_{w}(x,z,\omega )$ in our example --- at the point $x=0$ in the
basis of the eigenmodes of the complete system, $H_{i}(x,z,\omega )$, see
Fig.~\ref{FullModes}. In other words, we calculate what energy each mode
would carry through the structure. Applying this expansion we obtain the
field evolution in an arbitrary point $x$ along the structure:
\begin{equation}
\begin{split}
H(x,z,\omega )& =\sum_{i}C_{i}(\omega )\cdot H_{i}(x,z,\omega ), \\
C_{i}(\omega )=\int H_{w}(& 0,z,\omega )\frac{k_{xi}(\omega )}{k(\omega
)\varepsilon (z,\omega )}H_{i}^{\ast }(0,z,\omega )dz,
\end{split}
\label{FullModesEQ1}
\end{equation}
where $H_{i}$ are the eigenmodes of the complete system, $H_{w}$ is the
input fundamental waveguide mode of the dielectric waveguide, $k_{xi}$ are
the wavevectors of the eigenmodes of the complete system. Using Eq.~%
\eqref{FullModesEQ1} it is easy to calculate the transmittance through the
structure, Fig.~\ref{tOmega2a}. The comparison of the direct numeric
simulations and the modes expansion method shows an excellent agreement as
it can be seen in\ Fig.~\ref{tOmega2a}.

\section{Conclusion}

We have examined the properties of a multi-layer hybrid device combining the
waveguide and surface plasmon polariton modes in a planar geometry. Our main
focus here was on the analysis of the feasibility to control mode coupling
efficiency through the system design rather than via using additionally
written gratings. We have demonstrated that by the proper design of the
multi-layer structures it is feasible to achieve substantial reduction in
the length of the device.

We have shown that in the considered scheme (without using additional
gratings) of the SPR-based sensors there exists an optimal length of the
metal film. Such an optimal length is defined by the characteristic
(coupling) distance of the energy exchange between the waveguide core and
SPP mode. Moreover, there is a set of preferable metal film lengths (sensing
region lengths), which are multiples of the coupling distance. The minimal
sensing region length offers advantage of higher sensitivity and better
signal-to-noise ratio and an improved resolution as it was observed in the
recent work \cite{Dwivedi_2008}. Minimization of the sensing region length
also provides a better contrast between resonant and non-resonant
transmission (transmission at resonant and non-resonant frequency), due to
the small dissipation losses in the shorter metal film. In general, SPR
sensor with the minimal metal film length utilizes the effect of SPP-photon
transformation rather then SPP dissipation in the metal plate. In other
terms our design analysis provide a guidance where to cut a metal film to
make a short sensing device. The obtained results might be important for
miniaturization and integration of a range of nano-optic systems such as
e.g. bio- and chemical sensors, frequency-selective polarizers, and other
devices.

We have shown that an appropriate choice of the operational (resonance)
frequency could be derived from a rather simple examination of the
intersections of the waveguide and SPP dispersion curves resulting in their
hybridization. This opens a way of an efficient excitation of the surface
plasmon resonance by the waveguide mode and vice versa. In the situations
when the required operational frequency (wavelength) is defined by the
external conditions our results can be used as the design rules for a
specific multi-layer device to operate at the desirable frequency.
Evidently, hybridization --- coupling of the waveguide and SPP modes occurs
not only at a single frequency point, but rather in some spectral interval.
Shift of an operational frequency from an optimal value leads to gradual
reduction of the efficiency. Our results indicate that a characteristic
length of the devices satisfying the resonance conditions (accompanied by a
high level of hybridization) required for an efficient energy transfer
between waveguide and SPP modes could be as short as tens of micrometers.
This is quite an important departure from the characteristic
millimeter-scale \cite{Harris_SA_1995,Harris_BB_1999,Ctyroky_SA_1999}
lengths of conventional devices. Our scheme can be potentially generalized
to other geometries, for instance, it can be applied to the cylindrical
(fiber) geometry. By an appropriate selection of the parameters the
operational (resonance) frequency could be shifted into the near infrared
region including telecom windows near $1500$~$\mathrm{nm.}$ This can be
achieved, for instance, by increasing the waveguide layer thickness up to $%
350~\mathrm{nm}$ and simultaneous decrease of the metal layer thickness down
to $30$~nm. The minimum transmission of the multi-layer structure considered
here depends essentially on the dielectric properties of the media adjacent
to the metal film (the superstrate). Therefore, small variations in the
refractive index of this medium would result in a strong shift of the
transmission dip offering new opportunities and perspectives for the sensor
applications.

\section*{Acknowledgment}

The authors would like to thank Dr. A.~Nikitin, Dr. T. Allsop and Dr.
S.~Derevyanko for helpful discussions. This work was supported by the Royal
Society, the European Programme INTAS (INTAS YS 05-109-5182) and STCU grant
No~3979.

\end{document}